\documentclass[11pt]{article}
\usepackage{amsmath,amssymb,color,graphics,epsfig}
\usepackage{multirow}
\usepackage{cite}

\usepackage{amsfonts,bbm}

\newcommand{\hoch}[1]{$\, ^{#1}$}

\makeatletter
\@addtoreset{equation}{section}
\makeatother

\newcommand{\be}{\begin{equation}}
\newcommand{\ee}{\end{equation}}
\newcommand{\bea}{\setlength\arraycolsep{2pt} \begin{eqnarray}}
\newcommand{\eea}{\end{eqnarray}}
\newcommand{\nn}{\nonumber}

\def\0{{\sst{(0)}}}
\def\1{{\sst{(1)}}}
\def\2{{\sst{(2)}}}
\def\3{{\sst{(3)}}}
\def\4{{\sst{(4)}}}
\def\5{{\sst{(5)}}}
\def\6{{\sst{(6)}}}
\def\7{{\sst{(7)}}}
\def\8{{\sst{(8)}}}
\def\sst#1{{\scriptscriptstyle #1}}

\begin{document}

\begin{flushright}
\end{flushright}

\vspace{25pt}
\begin{center}
{\large {\bf Shadow and Quasi-Normal Modes of Schwarzschild-Hernquist Black Hole}}

\vspace{40pt}
{\bf Xing-Hui Feng\hoch{*} and Guang-Yu Zhang}

\vspace{10pt}

{\it Center for Joint Quantum Studies and Department of Physics,
School of Science, Tianjin University, Tianjin 300350, China}

\vspace{40pt}

\underline{ABSTRACT}
\end{center}

In this paper we study the shadow and quasi-normal modes (QNMs) of a black hole (BH) surrounded by a dark matter halo with Hernquist-type density distribution, which was reported in Ref. \cite{Cardoso:2021wlq}. In astrophysical scenarios, we find that the shadow radius enlarges as the compactness of halo increases. Therefore, we obtain an upper bound for the compactness ${\cal C}\le0.092$ with the Event Horizon Telescope (EHT) observations. We calculate axial gravitational QNMs of the galactic BH up to ${\cal C}\sim{\cal O}(1)$, and fit the redshift relative to Schwarzschild QNMs up to second order in the compactness (for ${\cal C}\le0.3)$. These highly redshifted QNMs, resulting from large compactness, are key to modeling the dark matter halo.

\vfill {\footnotesize \hoch{*} xhfeng@tju.edu.cn}

\thispagestyle{empty}

\pagebreak

\addtocontents{toc}{\protect\setcounter{tocdepth}{2}}


\vspace{1cm}

\section{Introduction}
Black holes (BHs) are a fundamental prediction of general relativity (GR). As cornerstone objects within GR, BH dynamics are significant for theory and observations, as extensively discussed in \cite{Chandrasekhar:1985kt}. Recent advancements, such as the images of the supermassive BHs at the center of M87$^\ast$ and SgrA$^\ast$ captured by the Event Horizon Telescope (EHT) and the detection of gravitational waves (GWs) from binary BHs and neutron stars by LIGO/Virgo, have ushered in a new era of BH astronomy. These groundbreaking experiments not only confirm the existence of BHs but also provide stringent observational constraints on theories of gravity in the strong-field regime.

It's unlikely that BHs are completely isolated objects in astrophysical scenarios. In fact, the aforementioned detections rely on interactions between BHs and their surroundings. BHs in the centers of galaxies are invariably surrounded by various distributions of matter, such as accretion disks \cite{accretion,Speri:2022upm} and dark matter halos \cite{Navarro:1995iw,Borriello:2000rv,Prada:2005mx}. The influences of dark matters on measurements of BH shadows and GW observations has been studied over the past decades. Most of these studies are phenomenological, where a halo is essentially put by hand by imposing a post-Newtonian potential onto the vacuum metric \cite{Macedo:2013qea,Cardoso:2019upw,Polcar:2022bwv,Vagnozzi:2022moj,Zhao:2023itk}. Recently, by employing "Einstein cluster" scheme, an exact spherically-symmetric BH immersed in a dark matter halo with Hernquist-type density distribution was obtained in \cite{Cardoso:2021wlq}. It is a fully-relativistic BH solution of the Einstein field equations surrounded by an anisotropic generic fluid. Subsequently, more self-consistent galactic BHs embedded in dark matter halos with different density profiles have been constructed analytically or numerically \cite{Jusufi:2022jxu,Konoplya:2022hbl,Daghigh:2023ixh,Figueiredo:2023gas,Speeney:2024mas,Shen:2023erj,Shen:2024qbb,Ovgun:2025bol,Fernandes:2025osu}, and their astrophysical implications have been investigated in \cite{Daghigh:2023ixh,Figueiredo:2023gas,Speeney:2024mas,Konoplya:2021ube,Cardoso:2022whc,Xavier:2023exm,Myung:2024tkz,Macedo:2024qky,Mollicone:2024lxy,Spieksma:2024voy,Pezzella:2024tkf,Chakraborty:2024gcr,Chowdhury:2025tpt,Gliorio:2025cbh,Datta:2025ruh,Alnasheet:2025tpd,Kouniatalis:2025itj,Destounis:2025tjn}.

Although the shadow and quasi-normal modes (QNMs) of the exact BH with a dark matter halo derived in \cite{Cardoso:2021wlq}, have been studied to some extent \cite{Konoplya:2021ube,Xavier:2023exm,Myung:2024tkz,Macedo:2024qky,Pezzella:2024tkf,Chakraborty:2024gcr}, a comprehensive understanding of their observational signatures is still lacking. This is because most results are confined to a limited parameter space, particularly focusing on extremely low compactness of the halo mass distribution. In this work, we give an intensive studies about these subjects across a broader parameter space. The rest of this paper is organized as follows. In section \ref{bhsolution}, we briefly review the construction and structure of BH with dark matter halo. Then we discuss its photon sphere and shadow in section \ref{shadow}. In section \ref{qnms}, we calculate the axial gravitational QNMs using three independent methods; the numerical results are presented in section \ref{numerical}. Finally, we conclude with a brief discussion in section \ref{discussion}.

\section{Schwarzschild-Hernquist black hole}\label{bhsolution}
To describe the spacetime geometry immersed in dark matter halo, it's convenient to take the following metric ansatz
\be
ds^2=-f(r)dt^2+\frac{dr^2}{1-2m(r)/r}+r^2d\Omega^2
\ee
where $d\Omega^2$ stands for the metric on a unit two-sphere. The "Einstein cluster" scheme leads to assuming an anisotropic fluid with vanishing radial pressure, such that \cite{Cardoso:2021wlq}
\be
T^\mu{}_\nu={\rm diag}(-\rho,0,P_t,P_t)
\ee
where $\rho$ is the density profile of the dark matter halo and $P_t$ is the tangential pressure. The Einstein equations $G_{\mu\nu}=8\pi T_{\mu\nu}$ give a set of equations of motion as follows
\bea
m'&=&4\pi r^2\rho\label{eqmass}\\
\frac{f'}{f}&=&\frac{2m}{r(r-2m)}\label{eqlapse}
\eea
In Ref. \cite{Cardoso:2021wlq}, the Hernquist-type density profile was considered
\be
\rho=\frac{M_{\rm DM}(a_0+2M_{\rm BH})(1-2M_{\rm BH}/r)}{2\pi r(r+a_0)^3}
\ee
where $M_{\rm DM}$ is the total mass of dark matter halo and $a_0$ is its characteristic scale. Note that to model the dark matter spike around the BH, the density profile is scaled with a factor $1-2M_{\rm BH}/r$, where $M_{\rm BH}$ is the mass of the central BH. The mass function can be obtained from \eqref{eqmass}
\be
m(r)=M_{\rm BH}+\frac{M_{\rm DM}r^2}{(a_0+r)^2}\left(1-\frac{2M_{\rm BH}}{r}\right)^2
\ee
The lapse function obtained from \eqref{eqlapse} is
\bea
f(r)&=&\left(1-\frac{2M_{\rm BH}}{r}\right)e^\Upsilon\\
\Upsilon&=&-\pi\sqrt{\frac{M_{\rm DM}}{\xi}}+2\sqrt{\frac{M_{\rm DM}}{\xi}}\arctan\frac{r+a_0-M_{\rm DM}}{\sqrt{M_{\rm DM}\xi}}\\
\xi&=&2a_0-M_{\rm DM}+4M_{\rm BH}
\eea
One can regard $e^{\Upsilon}$ as a redshift factor. At asymptotic infinity, $f(r)$ behaves
\be
f(r)=1-\frac{2(M_{\rm BH}+M_{\rm DM})}{r}+{\cal O}\left(\frac{1}{r^3}\right)
\ee
So the ADM mass of spacetime is $M_{\rm ADM}=M_{\rm BH}+M_{\rm DM}$. Noth that the event horizon is located at $r_h=2M_{\rm BH}$ as for the Schwarzschild solution. This solution could be regarded as a model for a supermassive BH in the center of a galaxy surrounded by a dark matter halo. For convenience to present, we refer it as Schwarzschild-Hernquist BH. To mimic astrophysical observation, one requires a hierarchy of scales: $M_{\rm BH}\ll M_{\rm DM}\ll a_0$. It's convenient to introduce a compactness parameter
\be
{\cal C}=\frac{M_{\rm DM}}{a_0}
\ee
to quantity the compactness of the dark matter halo. Usually the galactic dynamics bounds ${\cal C}\ge10^{-4}$ \cite{Navarro:1995iw}. To facilitate the analysis of the observational properties of Schwarschild-Hernquist BH, we define another parameter
\be
\quad \epsilon=\frac{M_{\rm BH}}{M_{\rm DM}}
\ee
which could be called the mass ratio. The hierarchy of scales means ${\cal C}\rightarrow0$ and $\epsilon\rightarrow0$. This is also the smooth limit to Schwarschild BH. In this work, we loosen such assumption and alow wide ranges of these two parameters.

\section{Photon sphere and shadow}\label{shadow}
\begin{figure}[t]
\center
  \includegraphics[width=5cm]{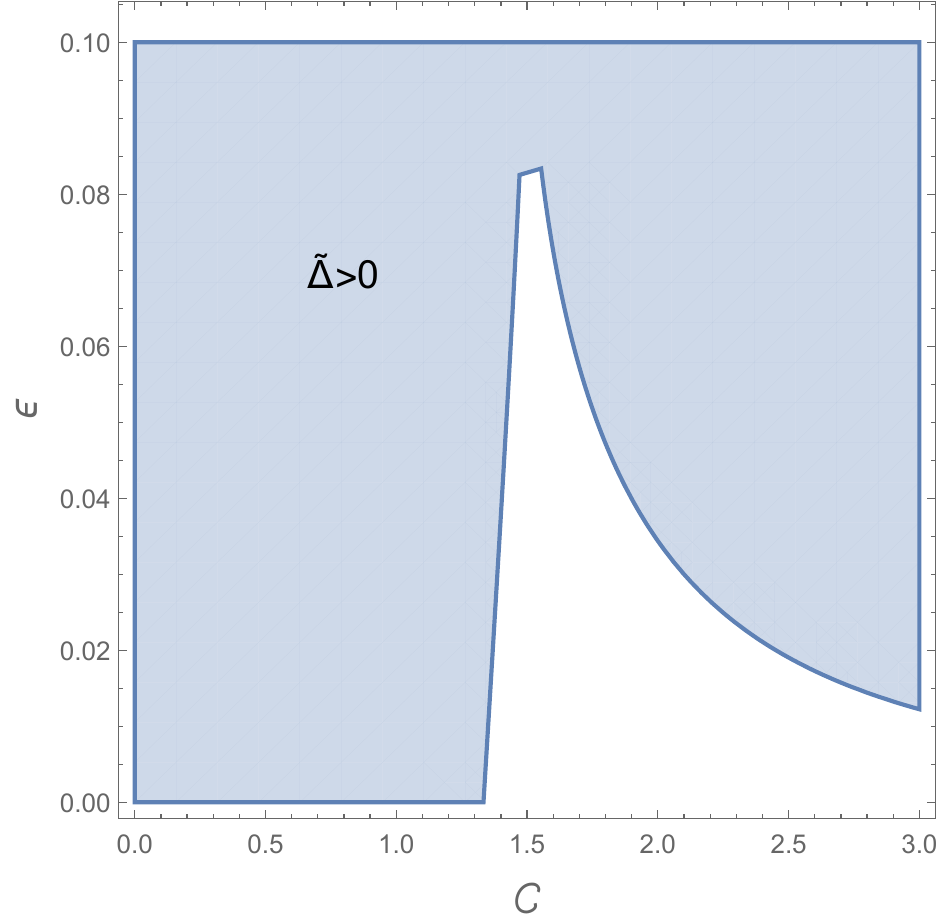}\qquad\qquad
  \includegraphics[width=7.5cm]{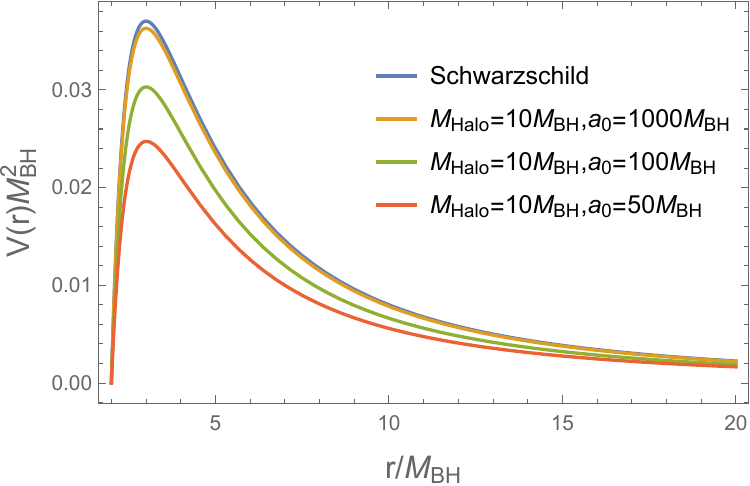}
  \includegraphics[width=6cm]{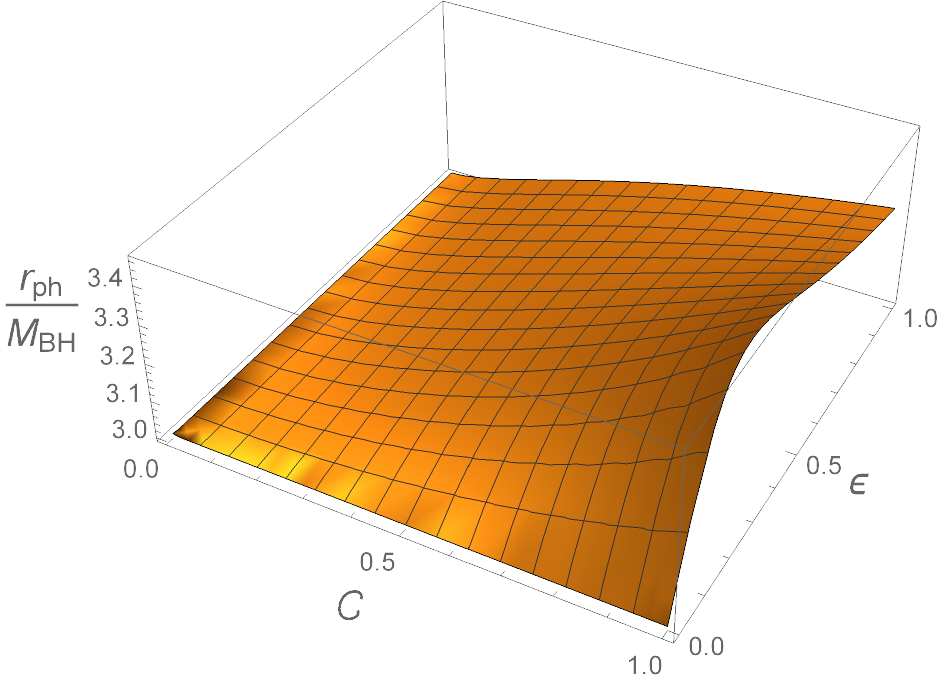}\qquad
  \includegraphics[width=7.5cm]{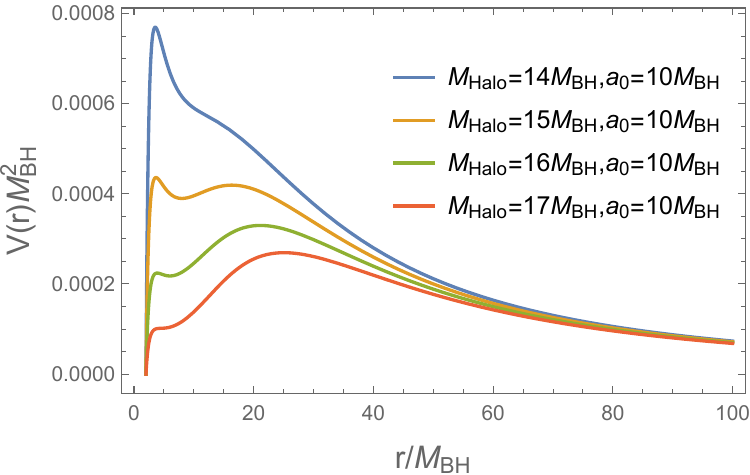}\\
  \caption{The top panel in the left column is the region (blue) for $\tilde\Delta>0$ in parameter space $({\cal C}, \epsilon$). The bottom panel in the left column is photon sphere $r_{ph}$ for various parameter $({\cal C}, \epsilon)$. The right column is the plot of effective potential $V_L$ of null geodesic.}\label{deltarph}
\end{figure}
The effective potential of null geodesic is
\be
V_L=\frac{f(r)}{r^2}
\ee
The photon sphere is determined by $V_L'(r)=0$, which is equivalent to $r-3m(r)=0$. This is a cubic order algebraic equation
\be
r^3+Ar^2+Br+C=0\label{cubic}
\ee
with $A=2a_0-3(M_{\rm BH}+M_{\rm DM}), B=a_0^2-6a_0M_{\rm BH}+12M_{\rm BH}M_{\rm DM}, C=-3M_{\rm BH}(a_0^2+4M_{\rm BH}M_{\rm DM})$.
The discriminant of \eqref{cubic} is $\Delta=\frac{p^3}{27}+\frac{q^2}{4}$, where $p=B-\frac{A^2}{3}$ and $q=\frac{2A^3}{27}-\frac{AB}{3}+C$. We have one real root when $\Delta>0$, while three real root when $\Delta<0$. In order to discuss the numbers of real root in various parameter space, we find the discriminant $\Delta$ gives
\be
\tilde\Delta=36 \epsilon \left(3 \epsilon^2-3 \epsilon+1\right) {\cal C}^3+36 \epsilon (3 \epsilon-2) {\cal C}^2+(36 \epsilon-3) {\cal C}+4
\ee
In the extremal mass ratio limit $\epsilon\rightarrow0$, i.e. $M_{\rm DM}\gg M_{\rm BH}$, $\tilde\Delta=4-3{\cal C}$. We plot the region for $\tilde\Delta>0$ in Figure \ref{deltarph}. We can see from the figure that an additional unstable photon sphere $r_{ph}$ is possible when the compactness ${\cal C}>4/3$. We cross check these results by plotting the effective potential $V_L$ in Figure \ref{deltarph}.

\begin{figure}[t]
\center
  \includegraphics[width=8cm]{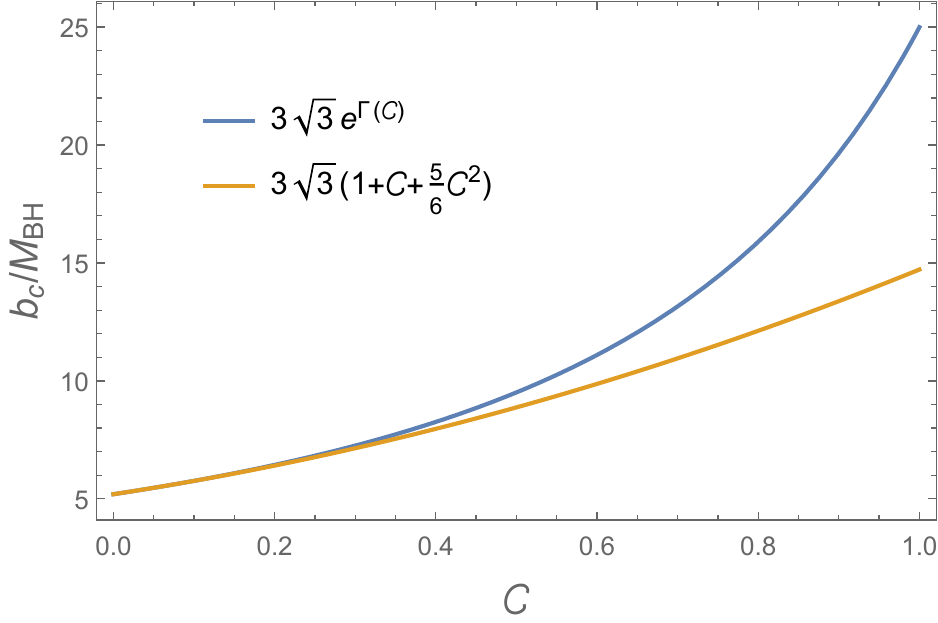}\\
  \caption{The critical impact parameter $b_c$ as a function of the compactness ${\cal C}$ when the mass ratio $\epsilon\rightarrow0$.}\label{bcC}
\end{figure}

In this paper we focus on only one photon sphere in relative low compactness. So we limit the compactness ${\cal C}<1$, and plot the photon sphere $r_{ph}$ in Figure \ref{deltarph}. The analytic express of $r_{ph}$ is very complicated whose form can be obtained according to appendix \ref{rootscubic}. Anyway we can take two limits, small compactness ${\cal C}$ or small mass ratio $\epsilon$. When the compactness is very low, i.e. ${{\cal C}\ll1}$, the photon sphere can be approximated as \cite{Cardoso:2021wlq}
\be
r_{ph}=3M_{\rm BH}(1+\epsilon{\cal C}^2)+{\cal O}({\cal  C}^3)\label{rph}
\ee
The corresponding critical impact parameter $b_c=r_{ph}/\sqrt{f(r_{ph})}$ is given by
\be
b_c=3\sqrt3M_{\rm BH}\left(1+{\cal C}+\frac{5-18\epsilon}{6}{\cal C}^2\right)+{\cal O}({\cal  C}^3)\label{bclowC}
\ee
When the mass ratio is very small, i.e. $\epsilon\ll1$, the photon sphere has the same expansion \eqref{rph}, while the critical impact parameter is
\be
b_c=3\sqrt3e^{\Gamma({\cal C})}M_{\rm BH}+{\cal O}(\epsilon),\quad \Gamma({\cal C})=\sqrt{\frac{{\cal C}}{2-{\cal C}}}\left[\frac{\pi}{2}+\arctan\left(\frac{{\cal C}-1}{\sqrt{{\cal C}(2-{\cal C})}}\right)\right]\label{bclargeC}
\ee
Note that this result is also valid for large compactness ${\cal C}\ge1$. It's easy to check that \eqref{bclargeC} reduces to \eqref{bclowC} when ${\cal C}\ll1$. We plot the critical impact parameter $b_c$ as a function of the compactness ${\cal C}$ in the extremal mass ratio $\epsilon\rightarrow0$ in Figure \ref{bcC}. We can see from the figure that the low compactness approximation is very accurate in the range ${\cal C}\le0.3$.

\section{Axial gravitational QNMs}\label{qnms}
QNM spectrum of black holes characterize response to fluctuations. In spherically-symmetric backgrounds, the perturbations are decomposed in terms of axial and polar tensor harmonics. In this work we focus on axial type perturbations, for which metric and matter fluctuations decouple. Then the axial gravitational perturbations are completely governed by a Schrodinger-like master wave equation \cite{Figueiredo:2023gas,Cardoso:2022whc,Feng:2024ygo}
\be
\frac{d^2\Psi}{dr_\ast^2}+(\omega^2-V)\Psi=0\label{master}
\ee
where $r_\ast$ is the tortoise coordinate, defined as
\be
dr_\ast=\frac{dr}{\sqrt{f(r)(1-2m(r)/r)}}
\ee
The effective potential $V$ reads
\be
V=\frac{f}{r^2}\left[\ell(\ell+1)-\frac{6m}{r}+m'\right]
\ee
when $m(r)=M_{\rm BH}$ we recover the usual vacuum Schwarzschild background and Eq. \eqref{master} reduces to the well-known Regge-Wheeler equation \cite{Regge:1957td}. By definition, QNMs satisfy the following boundary conditions,
\be
\Psi(r_\ast\rightarrow\infty)\propto e^{-i\omega r_\ast},\qquad \Psi(r_\ast\rightarrow-\infty)\propto e^{i\omega r_\ast}
\ee
which are requirement of the purely ingoing waves at the event horizon $(r_\ast\rightarrow-\infty)$ and purely outgoing wave at spatial infinity $(r_\ast\rightarrow\infty)$.

In this paper, we use three methods to calculate QNMs: matrix method, pseudospectral method and WKB method. Matrix and paseudospectral methods are very similar, both convert differential  equation \eqref{master} to an algebraic equation using discretization techniques. In order to implement matrix and pseudospectral method to equation \eqref{master}, we make the coordinate transformation
 \be
x=1-\frac{r_h}{r}
\ee
where $r_h=2M_{\rm BH}$ is the black hole horizon. We constraint our analysis to the outer region of the black hole, such that $r_h\le r<\infty$. Hence in terms of the new coordinate this region is bounded to the interval $x\in[0,1]$.

By taking into account pure ingoing waves at the event horizon and pure outgoing waves at spatial infinity, we have the boundary conditions
\bea
\phi(x)&=&x^{\frac{-i\omega r_h}{\sqrt{r_hf'(r_h)}}},\qquad x\rightarrow0\\
\phi(x)&=&e^{\frac{i\omega r_h}{1-x}}(1-x)^{-2i\omega M_{\rm ADM}},\qquad x\rightarrow1
\eea
We can assume the solutions which satisfy the boundary conditions as
\be
\phi(x)=e^{\frac{i\omega r_h}{1-x}}(1-x)^{-2i\omega M_{\rm ADM}}x^{\frac{-i\omega r_h}{\sqrt{r_hf'(r_h)}}}R(x)
\ee
In this case, the boundary conditions become
\be
R(0)=const.,\qquad R(1)=const.
\ee
To avoid numerical singularities at the boundaries, we can make an additional substitution
\be
\chi(x)=x(1-x)R(x)
\ee
This further simplify the boundaries as
\be
\chi(0)=\chi(1)=0
\ee
The final master wave equation can be expressed as
\be
A_2(x,\omega,\omega^2)\chi''(x)+A_1(x,\omega,\omega^2)\chi'(x)+A_0(x,\omega,\omega^2)\chi(x)=0\label{eqchi}
\ee

\subsection{Pseudospectral method}
The basic idea of pseudospectral method is to expand the regular function $\chi(x)$ in a base composed by cardinal functions $C_j(x)$, in the form
\be
\chi(x)=\sum_{j=0}^N g_jC_j(x)
\ee
The next step is to discretize the differential equation \eqref{eqchi} on a grid of collocation points. The best choice is the Gauss-Lobato grid given by
\be
x_i=\frac12\left(1-\cos\left[\frac{i}{N}\pi\right]\right),\quad i=0,1,2,\cdots,N
\ee
For this grid, it's natural to choose the Chebyshev polynomials $T_j(x)=\cos(j\arccos x)$ as the cardinal functions. Evaluating on the grid results in a matrix equation
\be
{\cal M}(\omega)g=({\tilde M}_0+{\tilde M}_1\omega+{\tilde M}_2\omega^2)g=0\label{quadratic}
\ee
where $g=(g_0,g_1,\cdots,g_N)^T$ and ${\tilde M}_i$ are numerical matrices of discretized coefficients. Now the solving of QNMs becomes a quadratic eigenvalue problem \eqref{quadratic}. A direct requirement to have non-trivial solution is
\be
\det({\cal M})=0\label{det}
\ee
The matrix equation \eqref{det} leads to an algebraic equation that depends on powers of $\omega$. We can solve it using build-in command {\it FindRoot} in {\it Mathematica}. The convergence of results need a small grid size at price of extremely long time consuming. Following \cite{Jansen:2017oag} we rewrite \eqref{quadratic} to obtain a linear form of the eigenvalue problem
\be
(M_0+M_1\omega)\vec{g}=0\label{linear}
\ee
where
\be
M_0=\left(
      \begin{array}{cc}
        {\tilde M}_0 & {\tilde M}_1 \\
        0 & \mathbbm1 \\
      \end{array}
    \right),\quad
M_1=\left(
      \begin{array}{cc}
        0 & {\tilde M}_2 \\
        -\mathbbm1 & 0 \\
      \end{array}
    \right),\quad
\vec g=\left(
  \begin{array}{c}
    g \\
    \omega g \\
  \end{array}
\right)
\ee
Then QNMs spectrum can be found by solving generalized eigenvalue problem \eqref{linear} via the {\it Eigenvalues} command in {\it Mathematica}. To avoid spurious eigenvalues we perform the calculations on two grids of different sizes and select only overlapping values . Note that the eigenvalues problem \eqref{linear} of calculating QNMs does not depend on any initial guess, as the secular equation \eqref{det}. Further the computing speed is rather high for small grid size due to involving only numerical matrices.

\subsection{Matrix method}
Matrix method is proposed in \cite{Lin,Lin:2016sch}. One first discretize variable $x\in[0,1]$ to $N$ points, from $x_1$ to $x_N$. Applying Taylor series around a reference point such as $x_2$, arbitrary univariate function $f(x)$ at each point and its derivatives at the reference point can be expressed as a matrix form
\be
\Delta{\cal F}={\cal M}D
\ee
where $\Delta{\cal F}$ is a $(N-1)\times1$ column vector,
\be
\Delta{\cal F}=(f(x_1)-f(x_2),f(x_3)-f(x_2),\cdots,f(x_j)-f(x_2),\cdots,f(x_N)-f(x_2))^T\\
\ee
while $D$ is a $N\times1$ column vector,
\be
D=(f'(x_2),f''(x_2),\cdots,f^{(j)}(x_2),\cdots,f^{(N)}(x_2))^T
\ee
and ${\cal M}$ is a $(N-1)\times N$ matrix,
\be
{\cal M}=\left(
           \begin{array}{cccccc}
              x_1-x_2 & \frac{(x_1-x_2)^2}{2} & \cdots & \frac{(x_1-x_2)^i}{i!} & \cdots & \frac{(x_1-x_2)^N}{N!} \\
              x_3-x_2 & \frac{(x_3-x_2)^2}{2} & \cdots & \frac{(x_3-x_2)^i}{i!} & \cdots & \frac{(x_3-x_2)^N}{N!} \\
              x_j-x_2 & \frac{(x_j-x_2)^2}{2} & \cdots & \frac{(x_j-x_2)^i}{i!} & \cdots & \frac{(x_j-x_2)^N}{N!} \\
              \vdots & \vdots & \vdots & \vdots & \ddots & \vdots \\
              x_N-x_2 & \frac{(x_N-x_2)^2}{2} & \cdots & \frac{(x_N-x_2)^i}{i!} & \cdots & \frac{(x_N-x_2)^N}{N!} \\
           \end{array}
         \right)
\ee
According to Cramer's rule, we can obtain $i$-th order derivative at the reference point
\be
D_i=\frac{\det({\cal M}_i)}{\det({\cal M})}
\ee
where ${\cal M}_i$ is the matrix formed by replacing the $i$-th column of ${\cal M}$ by the column vector $\Delta{\cal F}$. For instance, $f'(x_2)=\det({\cal M}_1)/\det({\cal M}), f''(x_2)=\det({\cal M}_2)/\det({\cal M})$. This way, we can express all the derivatives at each point as linear combinations of the function values $f(x_i)$. Substituting the derivatives into the eigenequation in study, one obtains $N$ equations with $f(x_1), f(x_2), \cdots , f(x_N )$ as its variables.

Implementing matrix discretization to Eq. \eqref{eqchi}, we obtain a matrix equation
\be
{\cal M}(\omega){\cal F}=({\tilde M}_0+{\tilde M}_1\omega+{\tilde M}_2\omega^2){\cal F}=0\label{quadratic}
\ee
where ${\cal F}=[\chi(x_1),\chi(x_2),\cdots,\chi(x_N)]^T$ is the vector of function values at grid points and ${\tilde M}_i$ are numerical matrices of discretized coefficients. Then we can use the same numerical technics as the pseudospectral method to find the QNMs.

\subsection{WKB method and eikonal limit}
The WKB approximation is an effective method to estimate QNMs with $\ell\ge n$. The general WKB formula can be written in the form of expansion around the maximum of the potential barrier \cite{Konoplya:2019hlu}
\bea
\omega^2&=&V_0+A_2(\nu^2)+A_4(\nu^2)+A_6(\nu^2)+\cdots\nn\\
&&-i\nu\sqrt{-2V_2}(1+A_3(\nu^2)+A_5(\nu^2)+A_7(\nu^2)+\cdots)
\eea
The QNMs boundary conditions means
\be
\nu=n+\frac12,\quad n=0,1,2\dots
\ee
where $n$ is the overtone number, and $V_i$ is the value of the $i$-th derivative of the effective potential at its maximum with respect to the tortoise coordinate. The functions $A_i$ for $i=2,3,4,\dots$ are the $i$-th WKB order correction terms to the eikonal limit, which depends on $\nu$ and derivatives of the potential at its maximum up to the order $2i$. the explicit forms of $A_i$ can be found in \cite{Iyer:1986np,Konoplya:2003ii,Konoplya:2004ip,Matyjasek:2017psv}. In this work we use the 6th order WKB method \cite{Konoplya:2003ii}.

We can simply discuss the corrections on QNMs with dark matter halo by considering the eikonal limit $\ell\gg n$. The light ring properties are connected QNMs in the eikonal regime. It was shown that the first order WKB formula gives an light ring/QNMs correspondence in the eikonal limit \cite{Cardoso:2008bp,Li:2021zct}
\be
\omega=\left(\ell+\frac12\right)\Omega-i\left(n+\frac12\right)\lambda\label{LRQNM}
\ee
where $\Omega$ is the angular velocity of the light ring and $\lambda$ is the Lyapunov exponent of the light ring. When $\ell$ is not very large, one can generalize the correspondence \eqref{LRQNM} to higher order eikonal approximations \cite{Konoplya:2023moy}. We assume the potential has a peak at
\be
V'(r_{max})=0
\ee
The point of maximum of the potential can be expanded in the eikonal limit as
\be
r_{max}=r_{ph}+\frac{r_2}{\kappa^2}+\frac{r_4}{\kappa^4}+\frac{r_6}{\kappa^6}+{\cal O}({\cal C}^3,\kappa^{-8})
\ee
where $\kappa=\ell+\frac12$ with expansion coefficients
\be
r_2= \left(1+4 \epsilon {\cal C}^2\right)M_{\rm BH},\quad r_4=\frac{(35+80 \epsilon {\cal C}^2)M_{\rm BH}}{12},\quad r_6=\frac{(1097-292 \epsilon {\cal C}^2)M_{\rm BH}}{144}
\ee
The 7th order WKB formula gives the 6th order eikonal limit
\be
b_c\omega=\left(\kappa+\frac{c_1}{\kappa}+\frac{c_3}{\kappa^3}+\frac{c_5}{\kappa^5}\right)-i\nu\left(d_0+\frac{d_2}{\kappa^2}+\frac{d_4}{\kappa^4}+\frac{d_6}{\kappa^6}\right)\label{eikonal}
\ee
The coefficients $c_i, d_i$ are presented in appendix \ref{coefficients}.

\section{Numerical results}\label{numerical}
\begin{figure}[t!]
  \centering
  \includegraphics[width=6.8cm]{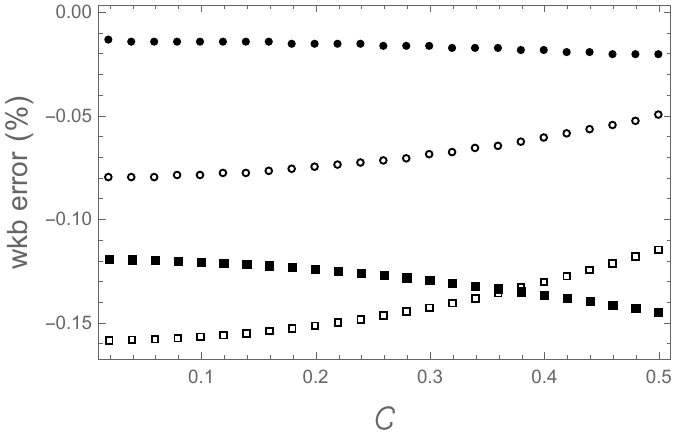}\qquad
  \includegraphics[width=6.6cm]{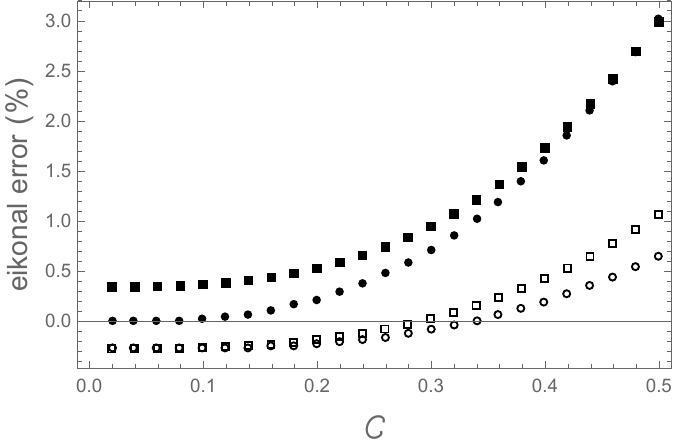}\\
  \includegraphics[width=7cm]{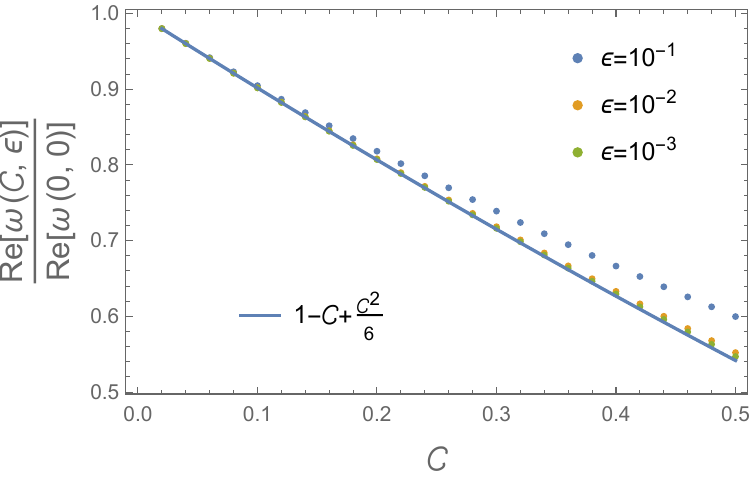}\quad
  \includegraphics[width=7cm]{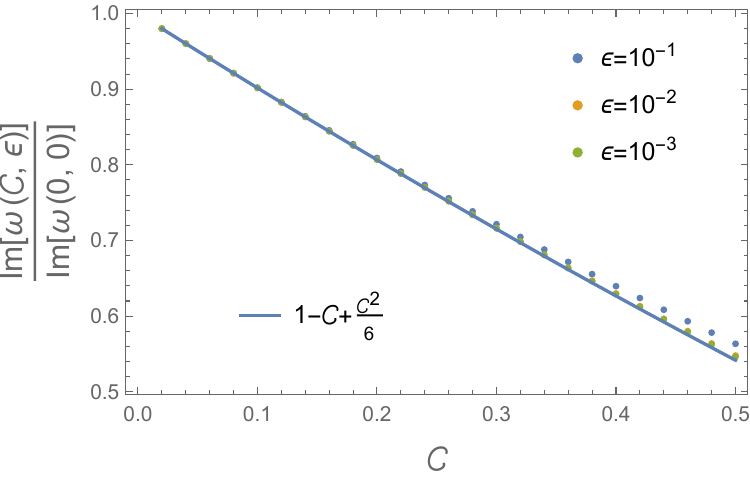}\\
  \caption{{\it Top panel:} Relative percentage difference between the $\ell=2$ QNMs with $\epsilon=0.1$ computed with the pseudospectral or matrix method and the 6th order WKB approximation (left), the 6th order eikonal limit (right). Filled (empty) markers correspond to the real (imaginary) parts, and circle (square) correspond to the $n=0$ fundmental ($n=1$ first overtone) QNMs. {\it Bottom panel:} Ratio between QNMs of Schwarzschild-Hernquist black hole and vacuum Schwarzschild black hole.}\label{error}
\end{figure}

In Tables \ref{qnml2} and \ref{qnml3} we show the numerical results of QNMs obtained by various methods. We can see that the pseudospectral and matrix methods match very well up to six digits of decimal. Note that for small $\epsilon$ we need very large $N$ to guarantee convergence, so we set $\epsilon=0.1$ as a typical parameters value. The 6th WKB method provides excellent agreement with numerical results for large $\ell$ and small $n$. We plot the relative errors between numerical results and WKB approximations (or eikonal limits) in Figure \ref{error}. We can see that the eikonal limit works bad when ${\cal C}>0.3$. We can obtain an elegant redshift formula of QNMs when $\epsilon$ approaches to zero according to the eikonal limit \eqref{eikonal}
\be
\frac{\omega({\cal C},\epsilon)}{\omega(0,0)}=\frac{3\sqrt3M_{\rm BH}}{b_c}=1-{\cal C}+\frac{{\cal C}^2}{6}+{\cal O}({\cal C}^3)
\ee
We also check this by plotting the numerical results in Figure \ref{error}. This form was obtained in \cite{Pezzella:2024tkf,Chakraborty:2024gcr} at leading order in the compactness (for ${\cal C}\le10^{-2}$).
\begin{table}[t!]
\center
\caption{Quasi-normal modes of the axial gravitational perturbations with $l=2$ and $\epsilon=0.1$.}\label{qnml2}
\begin{tabular}{c|c|c|c|c|c}
  \hline
  ${\cal C}$ & $n$ & Pseudospectral & Matrix & 6th WKB & 6th Eikonal \\ \hline\hline
  \multirow{2}{*}{0.1}
  & 0 & 0.338084-0.080250$i$ & 0.338084-0.080250$i$ & 0.338035-0.080187$i$ & 0.338159-0.080032$i$ \\
  & 1 & 0.313590-0.247123$i$ & 0.313590-0.247123$i$ & 0.313211-0.246735$i$ & 0.314739-0.246459$i$ \\ \hline
  \multirow{2}{*}{0.2}
  & 0 & 0.305715-0.071970$i$ & 0.305715-0.071970$i$ & 0.305668-0.071916$i$ & 0.306377-0.071807$i$ \\
  & 1 & 0.283342-0.221683$i$ & 0.283342-0.221683$i$ & 0.282990-0.221346$i$ & 0.284838-0.221265$i$ \\ \hline
  \multirow{2}{*}{0.3}
  & 0 & 0.276139-0.064176$i$ & 0.276139-0.064176$i$ & 0.276093-0.064132$i$ & 0.278094-0.064119$i$ \\
  & 1 & 0.255641-0.197732$i$ & 0.255641-0.197732$i$ & 0.255309-0.197450$i$ & 0.258061-0.197776$i$ \\
  \hline
\end{tabular}
\end{table}
\begin{table}[t!]
\center
\caption{Quasi-normal modes of the axial gravitational perturbations with $l=3$ and $\epsilon=0.1$.}\label{qnml3}
\begin{tabular}{c|c|c|c|c|c}
  \hline
  ${\cal C}$ & $n$ & Pseudospectral & Matrix & 6th WKB & 6th Eikonal \\ \hline\hline
  \multirow{2}{*}{0.1}
  & 0 & 0.542246-0.083612$i$ & 0.542246-0.083612$i$ & 0.542246-0.083612$i$ & 0.542394-0.083606$i$ \\
  & 1 & 0.526963-0.253733$i$ & 0.526963-0.253733$i$ & 0.526961-0.253726$i$ & 0.527157-0.253721$i$ \\ \hline
  \multirow{2}{*}{0.2}
  & 0 & 0.490059-0.074950$i$ & 0.490059-0.074950$i$ & 0.490059-0.074950$i$ & 0.491092-0.074985$i$ \\
  & 1 & 0.476052-0.227485$i$ & 0.476052-0.227485$i$ & 0.476050-0.227478$i$ & 0.477019-0.227633$i$ \\ \hline
  \multirow{2}{*}{0.3}
  & 0 & 0.442273-0.066782$i$ & 0.442273-0.066782$i$ & 0.442273-0.066782$i$ & 0.445267-0.066911$i$ \\
  & 1 & 0.429384-0.202730$i$ & 0.429384-0.202730$i$ & 0.429382-0.202724$i$ & 0.432090-0.203239$i$ \\
  \hline
\end{tabular}
\end{table}

\section{Discussion}\label{discussion}
We have studied the shadow and QNMs of BH immersed in a dark matter halo with Hernquist-type density profile, according to the solution reported in Ref. \cite{Cardoso:2021wlq}. It's obvious that the results depend on two parameters $(M_{\rm DM},a_0)$ of the halo. In order to analyze our results, we introduced two dimensionless parameters: compactness ${\cal C}$ and mass ratio $\epsilon$. For astrophysical scenarios with mass ratio $\epsilon\le10^{-2}$, the compactness ${\cal C}$ becomes the primary parameter of concern. We can constraint ${\cal C}$ using various observational data. For instance, the galactic dynamics bounds ${\cal C}\ge10^{-4}$ \cite{Navarro:1995iw}.

The EHT collaboration aims to image the central BH at M87$^\ast$ and SgrA$^\ast$. Given the current precision of the EHT, the errors associated with their results are around 10\%. This means that the relative deviation up to 10\% from the vacuum Schwarzschild/Kerr BHs cannot be distinguished with the current EHT data. For an observer located at infinity, the shadow radius of an asymptotically-flat and spherically-symmetric BH is described by the critical impact parameter $b_c$. If the relative deviation from Schwarzschild result $(b_c^{\rm Sch}=3\sqrt3M_{\rm BH}=5.196M_{\rm BH})$ is less than 10\%, i.e.
\be
4.677M_{\rm BH}\le b_c \le5.716M_{\rm BH}
\ee
the solution is in the favorable region. According to our result \eqref{bclargeC}, the upper bound of $b_c$ gives an upper bound on compactness ${\cal C}\le0.092$. A very close result can be found in \cite{Myung:2024tkz}.

The impact of astrophysical environments on GW observation is very important, since they can be thought of as a source of systematic errors in the program of testing GR. On the other hand, dark matter-dominated environment will display different GW signatures, which could help constrain the underlying model of dark matter halo. Recent research shows that the axial QNMs due to the environmental effects of dark matter halo can be viewed as overall redshift relative to vacuum Schwarzschild QNMs, and the redshift magnitude is proportional to the compactness (for ${\cal C}\le10^{-2})$ \cite{Pezzella:2024tkf,Chakraborty:2024gcr}. We calculated axial gravitational QNMs of Schwarzschild-Hernquist BH up to ${\cal C}\sim{\cal O}(1)$, and generalized the redshift to second order in compactness ${\cal C}$. The fitting formula is sufficiently accurate up to ${\cal C}\le0.3$. These highly redshifted QNMs due to large compactness are key to modeling the dark matter halo.

\section*{Acknowledgements}
This work is supported by NSFC (National Natural Science Foundation of China) Grant No. 11905157 and No. 11935009, Tianjin University Self-Innovation Fund Extreme Basic Research Project Grant No. 2025XJ21-0007.

\appendix
\section{The roots of cubic equation}\label{rootscubic}
The general cubic equation is
\be
y^3+ay^2+by+c=0
\ee
We can make a substitution $y=x-\frac{a}{3}$, then the equation becomes
\be
x^3+px+q=0\label{eqx}
\ee
with
\be
p=b-\frac{a^2}{3},\quad q=\frac{2a^3}{27}-\frac{ab}{3}+c
\ee
The discriminant of equation \eqref{eqx} is $\Delta=\frac{p^3}{27}+\frac{q^2}{4}$. When $\Delta<0$, the equation \eqref{eqx} exists three real roots. While $\Delta>0$, the equation exists only one real root. We focus on the $\Delta<0$ case, the real root is given by
\be
x=\sqrt[3]{-\frac{q}{2}+\sqrt\Delta}+\sqrt[3]{-\frac{q}{2}-\sqrt\Delta}
\ee

\section{Coefficients in the eikonal limit}\label{coefficients}
\begin{small}
\bea
c_1&=&\frac{-60 \nu ^2-547}{432} +\frac{\left(121-84 \nu ^2\right)\epsilon {\cal C}^2}{72}  \nn\\
c_3&=&\frac{854160 \nu ^4-8009976 \nu ^2-20776811}{40310784}+\frac{\left(339024 \nu ^4+3447624 \nu ^2+6655577\right) \epsilon {\cal C}^2}{3359232}\nn\\
c_5&=&\frac{596043168 \nu ^6+6244140960 \nu ^4-80576562918 \nu ^2-42878891327}{313456656384}\nn\\
&&+\frac{\left(729924000 \nu ^6-5652302112 \nu ^4+22273318266 \nu ^2+593091781\right) \epsilon {\cal C}^2}{5804752896}\nn\\
d_0&=&1-3 \epsilon {\cal C}^2,\quad d_2=\frac{940 \nu ^2-6599}{15552}+\frac{\left(916 \nu ^2+1207\right) \epsilon {\cal C}^2}{1728}\nn\\
d_4&=&\frac{-11273136 \nu ^4+258040200 \nu ^2-1541370007}{2902376448}+\frac{\left(89898096 \nu ^4-997116072 \nu ^2+6456421523\right) \epsilon {\cal C}^2}{967458816}\nn\\
d_6&=&\frac{-347667122880 \nu ^6+1720421667888 \nu ^4+11764433868044 \nu ^2+12361826419077}{135413275557888}\nn\\
&&+\frac{\left(-3856933570752 \nu ^6+34452468208560 \nu ^4-123263247487508 \nu ^2+752289005486157\right) \epsilon {\cal C}^2}{45137758519296}\nn\\
\eea
\end{small}

\end{document}